\documentclass[useAMS,usenatbib,usegraphicx]{mn2e}
\usepackage{amssymb}
\usepackage{amsmath}
\usepackage{times}
\usepackage[koi8-r]{inputenc}

\newcommand{\pasp}{PASP}
\newcommand{\apjs}{ApJS}
\newcommand{\apj}{ApJ}
\newcommand{\aaa}{A\&A}

\newcommand{\aj}{AJ}
\newcommand{\mnras}{MNRAS}

\title[ A new Luminous Variable in M33]{A new Luminous Variable in M33}
\author[Valeev et al.]{A.F. Valeev,\thanks{E-mail: 
azamat@sao.ru, olga@sao.ru, fabrika@sao.ru}
O. Sholukhova, and S. Fabrika
\\
Special Astrophysical Observatory, Nizhnij Arkhyz 369167, Russia}

\begin{document}

\date{\today}
\pagerange{\pageref{firstpage}--\pageref{lastpage}} \pubyear{2008}
\maketitle
\label{firstpage}

\begin{abstract}
We present a new 
luminous star in M\,33 located in the nuclear region. The star shows strong
Fe\,II and [Fe\,II] forest, hydrogen emissions in the spectrum,
as well as nebular lines. 
Ti\,II and Si\,II lines were detected in absorption, their radial velocity 
shifted by $\approx -30$\,km/s relative to emission lines. 
The star is variable over seven years with 0\fm5 variations
over a year. We studied its spectral energy distribution together with 
five confirmed Luminous Blue Variables and Var\,A in 
M\,33 using homogeneous data and methods. We found the star's bolometric luminosity 
to be log\,L/L$_\odot$\,$\approx6.27$, a surface temperature of T\,$\sim 16000$\,K and 
black body temperatures of two dust components of T\,$\sim 900$ and 420\,K.  
The new star has properties intermediate between the LBVs and Var\,A
(probable cool hypergiant). In the same time it has a hot photosphere, LBV-like 
luminosity and an extensive circumstellar material (strong [CaII] lines).
In these seven luminous variables in M\,33 we find the total range of the  
hot component luminosities is 1.0 dex, but that of the dust componets is 2.0 dex. 
We conclude that the dust phenomenon in the luminous variables is temporary and variable,   
and that dust activity may follow strong eruptions.

\end{abstract}

\maketitle

\section{Introduction}
\label{sec:intro}

Luminous Blue Variables (LBV) are the most massive stars in one of the final stages of evolution
\citep{HD1994_LBVreview}. They are very 
important for understanding of the late stages of evolution, production of WR stars, Supernovae, 
neutron stars and black holes. Their role in evolution of galaxies is significant.
Stellar evolution calculations \citep[e.g., ][]{Meynet2007} propose  
clear schemes of massive star evolution, however, the sequence of stellar 
conversions is not well understood yet \citep{Smith_Owocki06,ContiWNHstars,Smith08}. 
In order to understand the final evolution stages of 
massive stars, we need to discover and study more objects at the LBV and LBV--related stages. 

\citet{HD1994_LBVreview} reviewed LBV stars and their data known up to that time. In the galaxy M\,33 they 
selected four obvious LBV objects -- Var\,B, Var\,C, Var\,2, Var\,83, and one LBV candidate
V\,532 (GR\,290, "Romano's star", \citep{Romano78}). Spectroscopy has confirmed \citep{Fabrika2000} the
LBV status of V\,532. In photometric studies \citep{Kurtev_etal2001} its typical LBV behaviour
has been confirmed as well. In following papers \citep{Fabrika_etal2005,Viotti2006,Viotti2007} the star
has been studied in more detail. The star Var\,A in this galaxy had all the properties typical for
LBV stars \citep{HubbleSandage53}, it was one of the visually brightest stars in M\,33
with an intermediate F-type spectrum. However, the star shown M-type spectrum (unusual for 
classical LBVs) after its huge brightening in 1950--1953, thus it was classified
\citep{Humph_etal1987,Humphreys_etal2006,Viotti2006} as a cool hypergiant. 

Searches for LBV candidates in M\,33 have been done using various methods  \citep{Neeseetal91,Spiller1992,Calzetti1995,FabShol95,MasseyUIT,Shol_etal97,FabShol99,Shol_Fab2000,Corral_Herrero03}.
The main technique is to search for H$\alpha$-emitting sources coinciding with the galaxy 
stars. Selections of early-type stars were preferable, however, known LBV objects
show various observed colors \citep{Sterken_etal2008}. Some works were originally addressed
to a search for SS\,433-like objects in M\,33 \citep{Neeseetal91,Calzetti1995,FabShol95}.
SS\,433 has a spectrum very similar to that of a late WN star \citep{Fab04}. 

A new principal step has been done by \citet{Massey2006} in their CCD survey (UBVRI,
H$\alpha$, [S\,II], [O\,III]) of Local Group galaxies. The final catalog  
in M\,33 contains 146,622 stars up to 23rd magnitude with photometrical accuracy of 1--2~\%.
\citet{Massey2007_Ha} describe a list of H$\alpha$ emission-line stars in Local Group galaxies.
They focused on a search for new LBVs based on their spectroscopic similarity to 
known LBVs. Their number of known and suspected LBVs in M33 is 37. Some of their 
candidates coincide with previously described LBV candidates in M\,33
\citep{Corall1996,Shol_etal97,Shol_etal99,Shol_Fab2000,Fabrika2000}.

We performed photometry of point-like H$\alpha$ emitters in the M\,33 images of \citet{Massey2006} 
and selected stars with $V<18.5$ mag, $B-V < 0.4$ having H$\alpha$ excess. We did independent 
aperture photometry in the UBVRI and H$\alpha$ images for selected stars. Compared to  
automatic PSF photometry (which is the only possible way of dealing with a huge number of stars 
\citep{Massey2006}), aperture photometry is controlled by eye to avoid confusion
in crowded (broad-band) and complex (H$\alpha$) regions. In the selection we used the 
same method as discribed by \citet{Fabrika_etal1997}. With a distance modulus of 24.9 mag 
\citep{Bonanos06} our list \citep{Valeev_etal09} covers bright blue supergiants in the galaxy.
In our follow up spectroscopy of the list stars we found a new luminous variable star.
Here we report a spectral and photometrical study of the star and examine its spectral energy 
distribution together with all the confirmed LBVs and Var\,A in M\,33. 

\section{Observations}
\label{sec:obs}

For spectroscopy we used the SCORPIO focal reducer \citep{AfMois05} of the Russian 
6-m telescope.
We took spectra of the new star on November 18, 2007 with a spectral range 
and resolution of $\lambda\lambda 3700-7200$~\AA \, 
and 12 \AA, on January 7, 2008 ($\lambda\lambda 3900-5700$~\AA \, and 5 \AA), and on 
January 8, 2008 ($\lambda\lambda 5700-7400$~\AA \, and 5 \AA). The slit width was $1{\arcsec}$, 
and seeing $\approx 1.5{\arcsec}$. The spectra were reduced using standard procedures.
The accuracy of radial velocity calibration is about 10\,km/s.

\begin{figure}
\center{
\includegraphics[width=50mm,angle=0,trim=0mm 0mm 0mm 0mm]{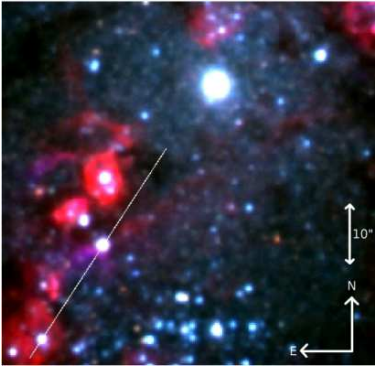}
}
\caption{
Nuclear region of M\,33, the bright blue object is the nucleus. The new star
(J2000, 013352.42+303909 or N\,93351 by \citet{Massey2006}) is indicated. The spectral slit
covering both the star and another blue candidate is shown. The image was obtained as rgb  
from H$\alpha$, V and [O\,III].
}
\label{image93351}
\end{figure}

In Fig.\,\ref{image93351} we show the central region of M\,33,
The star is located 34.3\arcsec  \, (160\,pc) from the nucleus to the
South-East. 
The faint H$\alpha$ filaments around the star may be related to it.
Our slit position was selected to cover the second blue star candidate
(Fig.\,\ref{image93351}).
In the HST archival broad band images the star is single and point--like.

For interstellar absorption estimates, both in this star and the other 
six stars in M\,33 (Var\,A, Var\,B, Var\,C, Var\,2, Var\,83 and V\,532),
we used H$\alpha$/H$\beta$ emission line ratios (Section\,\ref{sec:SEDs}) measured in 
surrounding nebulae. We used our own spectra of these targets taken with 
the same telescope and the MPFS instrument (IFU mode \citep{manualMPFS}) 
in December, 2004 and October, 2005 and with SCORPIO (December, 2004 and October, 2008). 
The spectral data were used for independent estimates of the star temperatures.

In studying the SEDs of the stars, we used published optical photometry 
(Section\,\ref{sec:SEDs}) and IR archival data from 2MASS and Spitzer. The 2MASS 
survey of M\,33 is available in J, H, K bands.
Spitzer observations of M\,33 
were made using all four bands of the 
Infrared Array Camera (3.6, 4.5, 5.8, and 8.0~\micron).
We used "post-basic calibrated data" and did photometry of all our targets
using the task {\it phot} in IRAF (http://ssc.spitzer.caltech.edu/irac/dh/).
All five Spitzer observations were averaged to have one photometrical measurement for each target.

\section{The new luminous variable N\,93351}
\label{sec:93351}
\begin{figure}
\vspace*{-0.3cm}
\includegraphics[width=\columnwidth,angle=0]{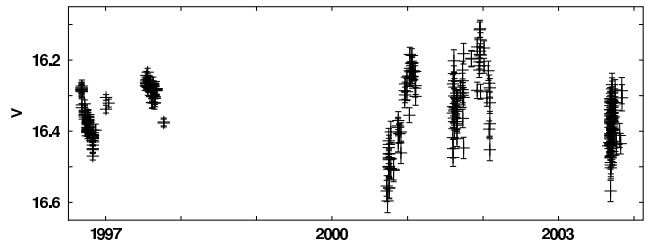}
\caption{
Light curve of N\,93351 in V-band. The data for 1997--1998 were taken from \citet{Direct2},
the data for 2000--2004 from \citet{Shporer2006}. 
A systematic shift -0.2 mag was applied to the second data. 
Photometrical errors are less than 0.03 mag.
}
\label{variability93351}
\end{figure}

The new star N\,93351 has a V magnitude of 16.16, and B-V=0.28, U-B=-0.44
\citep{Massey2006}. A light curve of the star 
is presented in Fig\,\ref{variability93351}. 
The star shows a variability for the past 7 years. It shows 0\fm5 variability
over a year. 

\begin{figure*}
\includegraphics[width=0.95\textwidth,angle=0]{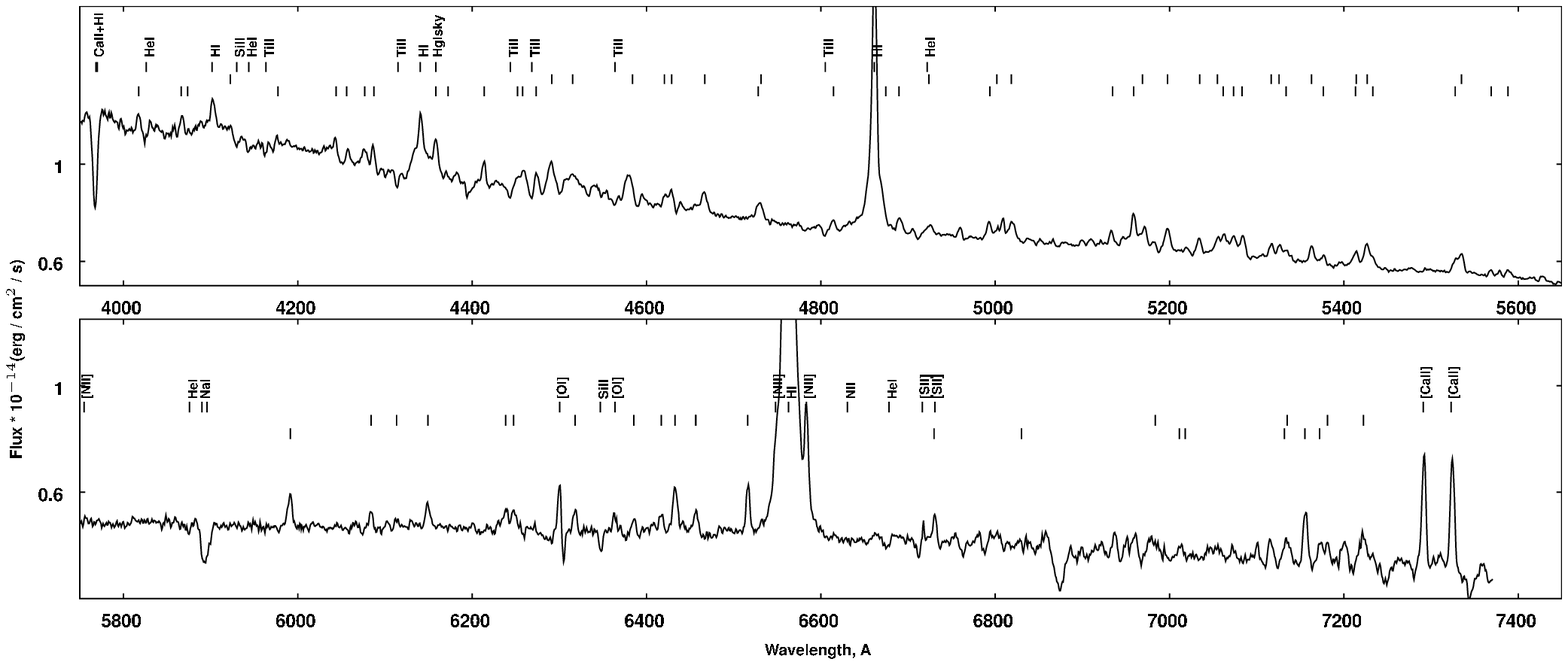}
\caption{
Spectra of N\,93351 in blue (top) and red (bottom) regions. Some lines are indicated (upper row),
the middle row of vertical bars marks FeII lines, the bottom row marks [Fe\,II] lines. CCD fringes 
are notable at wavelengths longer than $\sim 6800$~\AA. 
}
\label{spectr1200G}		
\end{figure*}

\begin{figure}
\includegraphics[width=60mm,angle=270]{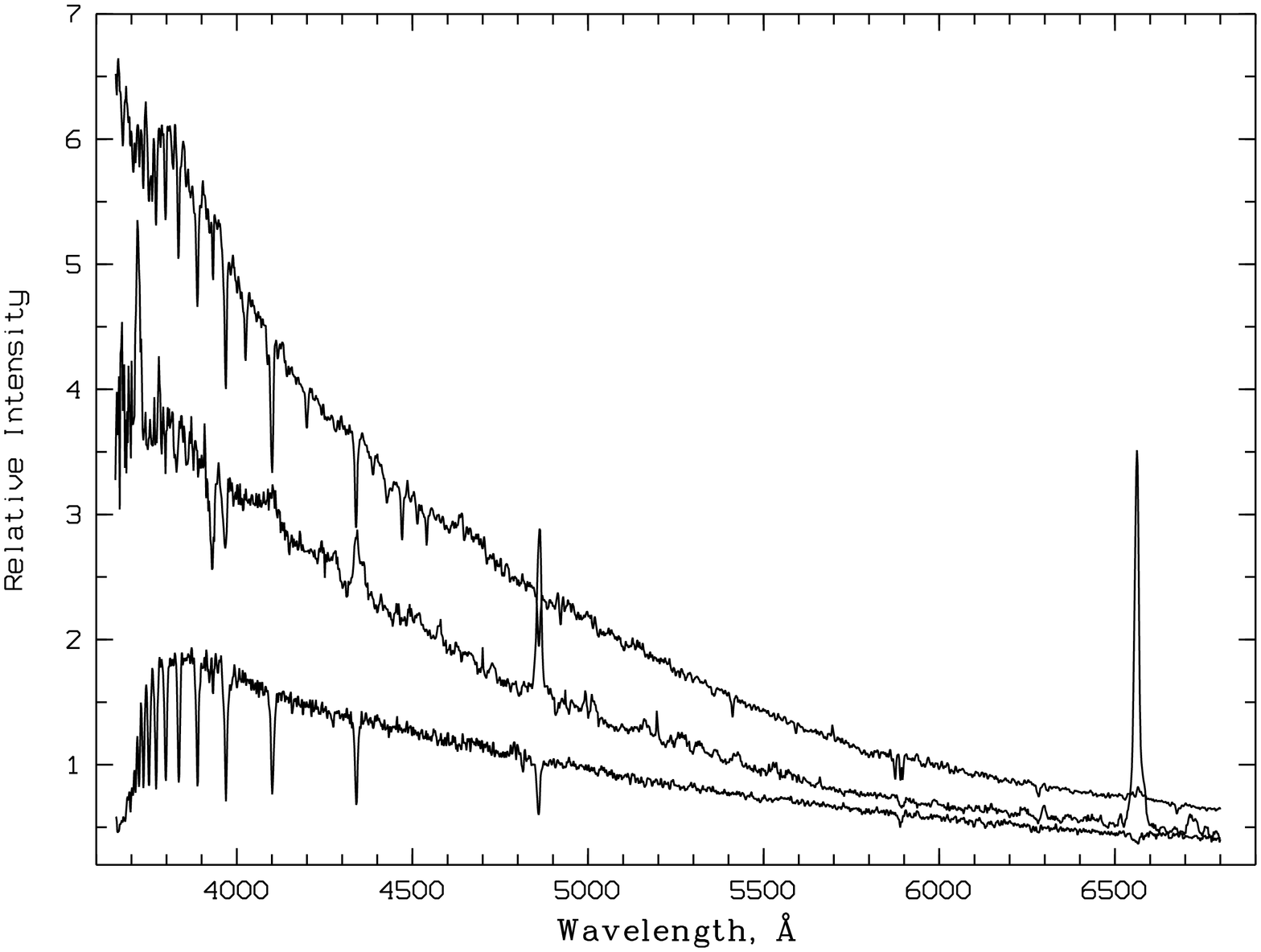}
\caption{
Dereddened spectra of N\,93351 (middle) together with standard stars O\,8I (HD\,225160, 
top) and B\,9Ib (LSI\,V\,P24, bottom) taken from \citet{Jacoby1984}. 
}
\label{spectr_comp}
\end{figure}

In Fig\,\ref{spectr1200G} we present spectrum of N\,93351 taken with a resolution 
of 5~\AA. The star has a blue continuum and very bright hydrogen emissions with 
broad wings.
The most remarkable feature is Fe\,II and [Fe\,II] emission forest. We detect He\,I,  
Ti\,II and Si\,II absorption lines. He\,I absorption may be partly filled with nebular 
emission, which is seen in the spectral image. 
We detect Na\,I\,D1,\,2 
and Ca\,II H,\,K lines. 
The last lines may be not only interstellar in origin. The spectrum presents 
strong [O\,II]\,$\lambda 3727$ line, Balmer absorption edge is practically absent
(Fig\,\ref{spectr_comp}). 
[O\,III]\,$\lambda \lambda4959, 5007$ lines are present, they are faint and 
crowded with bright Fe\,II emissions. There are collisionally excited nebular lines of 
[S\,II]\,$\lambda \lambda6717, 6731$, [N\,II]\,$\lambda \lambda6548, 6583$ (and 
marginally [N\,II]\,$\lambda5755$) in the spectrum.

The spectrum is quite typical for LBVs. 
However we observe very strong [Ca\,II]\,$\lambda \lambda  7291, 7323$ emissions, which 
are not present in classical LBVs. These lines are seen in the cool hypergiants
Var\,A in M\,33 and IRC+10420 \citep{Humph_etal1987,Humphreys_etal2006} with 
extensive circumstellar material. We conclude therefore, that N\,93351 has intermediate 
spectrum between classical hot LBV stars and the cool hypergiants like Var\,A. 

\begin{table}
\caption{Equivalent  widths, relative intensities and dereddened
($A_V=1.0)$ relative intensities in units of H$\beta$ intensity of
main not blended lines.
}
\label{lines}
\bigskip
\begin{tabular}{|l|l|l|l|}
\hline
Line                   & EW\,($\pm 1\sigma$), \AA & I(obs) & I(dered) \\
\hline
[OII]\,$\lambda3727$   & 4.9  (0.3)              & 0.63    & 0.86   \\
CaII\,K (abs)          & 2.8  (0.3)              &         &        \\
H$\delta$              & 1.0  (0.1)              & 0.11    & 0.13   \\

[FeII]\,$\lambda 4017$ & 0.30 (0.02)             & 0.03    & 0.04   \\

[FeII]\,$\lambda 4066$ & 0.33 (0.02)             & 0.04    & 0.05   \\
H$\gamma$              & 2.1  (0.2)              & 0.18    & 0.21   \\

[FeII]\,$\lambda 4413$ & 0.50 (0.02)             & 0.04    & 0.05   \\
FeII\,$\lambda 4666$   & 0.80 (0.03)             & 0.07    & 0.07   \\
H$\beta$               & 14.1 (0.1)              & 1.0     & 1.0   \\

[FeII]\,$\lambda5158$  & 1.1 (0.02)              & 0.07    & 0.07   \\
FeII\,$\lambda5197$    & 0.86 (0.03)             & 0.05    & 0.05   \\
FeII\,$\lambda5362$    & 0.73 (0.02)             & 0.04    & 0.04   \\
NaI\,D1,D2 (abs)       & 1.5  (0.3)              &         &       \\

[FeII]\,$\lambda5991$    & 1.70 (0.1)              & 0.08    & 0.06   \\
FeII\,$\lambda6149$    & 1.12 (0.05)             & 0.05    & 0.04   \\
FeII\,$\lambda6432$    & 1.86 (0.05)             & 0.09    & 0.06   \\
FeII\,$\lambda6516$    & 1.78 (0.05)             & 0.08    & 0.06   \\
H$\alpha$              & 124  (0.5)              & 5.60    & 3.98   \\

[SII]\,$\lambda 6717+6731$&9.0 (0.4)            & 0.23    & 0.16   \\

[FeII]\,$\lambda7155$  & 3.6 (0.1)               & 0.12    & 0.08   \\

[CaII]\,$\lambda7291$  & 9.0 (0.1)               & 0.27    & 0.18   \\

[CaII]\,$\lambda7323$  & 8.9 (0.1)               & 0.26    & 0.17   \\   

\hline
\end{tabular}
\end{table}

The radial velocity of Fe\,II and nebular lines is $\approx -120$\,km/s, which is 
in accordance with the H\,I velocity \citep{HI} in this area of the galaxy. 
We observe a blue shift of $\sim -30$\,km/s in all absorption lines relative to 
hydrogen emission lines.
This shift may arise 
in an expanding photosphere of the star.  
Fe\,II and [Fe\,II] line widths correspond to our spectral resolution (5\,\AA, measured in 
sky lines). For the H$\alpha$ and H$\beta$ line profiles we performed gaussian analysis 
for broad and narrow components and found the component widths 
(corrected for the instrumental resolution) of
800\,km/s and 180\,km/s in H$\alpha$, 950\,km/s and 110\,km/s in H$\beta$ correspondingly. 
Radial velocities of the narrow components in these two lines are about the same, 
$-106 \pm 5$\,km/s, and velocities of the broad components are  $-125 \pm 10$\,km/s.

Table\,\ref{lines} presents intensities and equivalent widths of main lines, 
which are not blended. The intensities are given in units of intensity of H$\beta$.
We present also the line intensities corrected for interstellar extinction, where 
we adopted $A_V=1.0$ (Section\,\ref{sec:SEDs}). To extimate the extinction we measured
H$\alpha$ and H$\beta$ line fluxes in the surrounding nebula at a distance from 
5$^{\prime\prime}$ to 13$^{\prime\prime}$ off the star along the slit. We found the 
fluxes per squared arcsecond F(H$\alpha) = (1.08 \pm 0.04) \cdot 10^{-14}$\,erg/cm$^2$ sec and 
F(H$\beta) = (2.5 \pm 0.1) \cdot 10^{-15}$\,erg/cm$^2$ sec. Corresponding line ratio
H$\alpha$/H$\beta$ = 4.0--4.5 gives $A_V = 1.0 - 1.4$. 
The equivalent width of Na\,I\,D\,1,\,2 absorption (Table\,\ref{lines}) confirms 
this value of extinction, a rough linear relation \citep{Barbon90} between the
doublet EW and reddening gives $A_V = 1.2 \pm 0.2$.
 
Fig.\,\ref{spectr_comp} presents spectrum of N\,93351 taken with resolution 12\,\AA \,
corrected for the extinction ($A_V=1.0$, E(B-V)=0.32) together with standard stars
O\,8I (E(B-V)=0.56) and B\,9Ib (E(B-V)=0.23), whose dereddened spectra were 
taken from \citet{Jacoby1984}. Both the spectral types and reddening values of the
standard stars are considered as approximate \citep{Jacoby1984}. 
Comparing the spectrum slope and intensity of the Balmer edge with those 
in classified stars we estimate roughly the temperature of N\,93351's 
photosphere of $13000 - 16000$\,K. Absorption Ca\,II\,K and Ti\,II lines may 
be formed in cooler extended atmosphere (the wind). 

\section{Spectral energy distributions} 
\label{sec:SEDs}

To estimate the temperature and luminosity of this star and other six M\,33 stars 
(Var\,A, Var\,B, Var\,C, Var\,2, Var\,83 and V\,532) 
we study their SEDs. 
Our main sources for the data are optical photometry by 
\citet{Massey2006} (October 2000 -- September 2001), 2MASS (December, 1997),
Spitzer (January 2004 -- June 2005) and our spectroscopy (December 2004 and October 2005). 
We use these data, otherwise describe individual cases. 
We checked all these stars in the literature for their variability so as not to use 
data related to different photometrical states.

We used our own spectra in order to have independent estimates of a reddening and 
a temperature of the stars. In the estimates of the reddening we used 
H$\alpha$/H$\beta$ flux ratios measured in nebulae surrounding the stars.
In the Case B of a standard
gaseous nebula this ratio is 2.87 (H$\gamma$/H$\beta = 0.47$, H$\delta$/H$\beta = 0.26$), 
and stays the same to within 10\,\% in a wide range of gas temperatures and densities 
\citep{Osterbrock_Ferl2006}. Obvious limitations of this method are that the nebula may
not be connected with the star, and the star may have some additional 
curcumstellar extinction. In two stars Var\,A and Var\,C there is no clear 
morphological association with the nebula. 

We used a standard method to fit observed 
SED components to black body (BB) spectra with a reddening (we adopt $R_V=3.1$).  
Using the best fit SED solutions we tried to satisfy the independently estimated
stellar temperatures (first) and the extinction values (second). We fitted the 
observed SEDs of the stars to BB spectra, suggesting that one hot spectrum is 
stellar and one or two cooler spectra for dust emission. 
We kept the temperature 
of cool components to less than 1500\,K. We estimated luminosities integrating 
the extinction corrected SED components. 

\begin{figure}
\includegraphics[angle=0,width=3.3in]{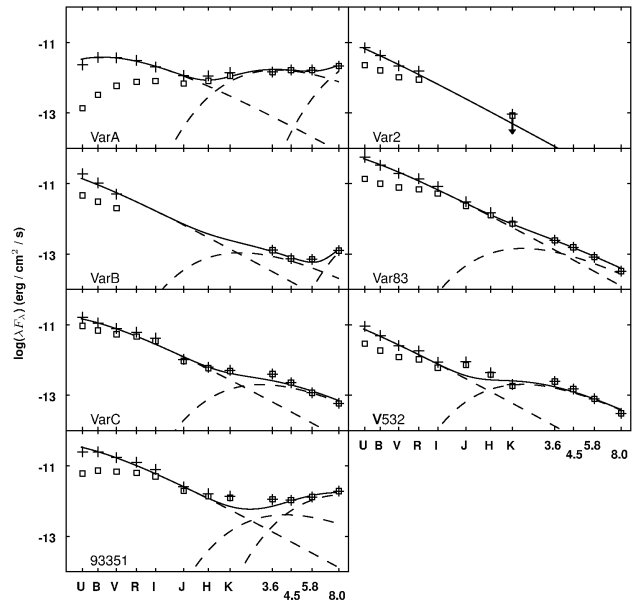}
\vspace{-0.3cm}
\caption{Spectral energy distributions of the studied stars in M\,33. There are observed data (squares),
data corrected for interstellar absorption (crosses), the best fit BB models of hot and dust 
components (dashed lines) and sum of the spectral components (solid lines).
}
\label{SED}
\end{figure}

In some cases we did not use U data, when this band fluxes were obviously weak 
(because of the Balmer absorption). As a rule we did not use R, I (and sometimes J) 
data, because they can contain a contribution of free-free emission from stellar winds. 
Fig.\,\ref{SED} presents the observed and intrinsic SEDs of the stars as well as the 
best fits of BB models. Derived parameters of the stars are presented in Table\,2
. 
Below we describe the individual stars in more detail.

\begin{table}
\label{starParams}
\caption{Parameters of the studied stars. Columns present the star name, adopted values of extinction, 
derived bolometric luminosity, temperature of the stars (T$_{s}$), and temperatures of warm (T$_{w}$) and cold dust (T$_{c}$) components. All temperatures are given in units of 1000\,K, lumimosities in solar units. 
Contributions of the dust components (per cent) to the total luminosities are given in brackets. 
The star (L$_{s}$) and dust  component (L$_{d}$) luminosities are given in last columns.
}
\bigskip
\begin{tabular}{|l|c|c|l|l|l|l|l|}
\hline
Name & $A_V$ & lg\,L & T$_{s}$ & T$_{w}$ & T$_{c}$ & lg\,L$_{s}$ & lg\,L$_{d}$\\
\hline
N\,93351 & 0.75&   6.13 &  13 &    0.7(3.4) &    0.19(26) & 5.98 & 5.60\\
         & 1.0 &   6.27 &  16 &    0.9(0.9) &    0.42(3.2) & 6.25 & 4.88 \\
VarA  &    2.0 &   5.63 &   8 &    1.0(15) &    0.22(50) & 5.17 & 5.44 \\
      &    2.5 &   5.98 &  10 &    0.90(7) &    0.18(64)  & 5.45 & 5.83 \\
VarB  &    1.0 &   6.21 &  25 &   1.5(0.3) &    0.18(2.5) & 6.20 & 4.66 \\
VarC  &    0.4 &   5.86 &  15 &   1.2(1.1) &    - &  5.86 & 3.90\\
Var2  &    0.8 &   6.34 &  40 & - & -  & 6.34 & - \\
Var83 &    0.7 &   6.32 &  18 &   1.5(0.3) &    -  & 6.32 & 3.80\\
      &    1.0 &   6.65 &  22 &   1.5(0.1) &    - & 6.65 & 3.65\\
V532  &    0.8 &   6.24 &  35 &   1.5(0.5) &    -  & 6.24 & 3.94\\
\hline
\end{tabular}
\end{table}

\noindent
{\bf N\,93351}. The star does not show strong variability.
Regarding the probable temperature and extinction ranges of the star found 
in Section~(\ref{sec:93351}), 
we find the best fit with $A_V \approx 1.0$ and T$\approx 16000$\,K (Fig.\,\ref{SED}). 
In Table\,2 
 we present a second fit version with $A_V = 0.75$ and T\,=\,12800\,K,
however we consider the higer temperature as more probable. The IR SED can not be fitted 
with one BB. We derive two dust components, with  T$\approx 900$\,K and T$\approx 420$\,K. 

\noindent
{\bf Var\,A}. 
In the period 1997 -- 2004 the star did not change its brightness notably 
\citep{Humphreys_etal2006}. From our spectrum we estimate a possible 
temperature range as $7000 <$ T $<15000$\,K. Actually there 
are only H\,I emissions and absorption blends of Fe\,II in the spectrum. 
A nearby nebula 
located in 9$^{\prime\prime}$-12$^{\prime\prime}$ to NE direction, gives 
$A_V=0.9-1.1$. 
This is the only star in our list whose luminosity was determined
quite firmly \citep{Humph_etal1987,Humphreys_etal2006}, lg\,L/L$_{\sun} 
\approx 6.0$ (with the distance to M\,33 adopted by us).  
These authors argue that the star has a notable amount of neutral
extinction. To be consistent we use the same method as we do for 
other stars, but we consider this analysis as formal. 
We find that the luminosity lg\,L/L$_{\sun} = 6.0$ may be reached
at $A_V=2.5$ (T=10000\,K). One needs better spectrum of Var\,A to find
correct extinction and the star temperature.


\noindent
{\bf Var\,B}.
The star shows strong variability. After its maximum of $V \sim 15$\,mag in 1992 
it weakened to 17.7 mag in 2005 \citep{Szeifert1996,Massey2006,Viotti2006,Shporer2006}. 
We used only UBV data from \citet{Viotti2006} and Spitzer data as about simultaneous. 
A surrounding nebula 
gives us the extinction limits $A_V=1.4-2.8$. We estimate  
the star's temperature range as 15000 -- 25000\,K (strong He\,I, [Fe\,II], [Fe\,III] lines, 
and no Bowen blend or He\,II).
Fitting the SED, we found the maximal value of $A_V$ we may adopt to be 1.0 mag 
(corresponding to a temperature of 25000\,K). Otherwise the temperature is too high. 
Note that in another brightness state of the star  
\citet{Szeifert1996} used $A_V=0.7$. They estimated extinction as foreground 
galactic plus 1/2 of the value estimated from H\,I measurements. 
In spite of the different methods the star luminosity they determined is the same as that we found.

\noindent
{\bf Var\,C}.
In 1985 the star showed a maximum brightness of B=15.4, and in 1997--2001 a minimum 
of V$\approx$17.3 with a variability of less than 0.2 mag, 
although it begun to brighten again until 2004 on $\approx 1$ mag 
\citep{Szeifert1996,Shporer2006}. 
We used our main sources of the data. 
A nebula measured to SE direction from the star to 11$^{\prime\prime}$ gives in
different parts $A_V$ from 0.32 to 0.73. The star temperature estimate is 10000 -- 15000\,K. 
There are Mg\,II lines, bright hydrogen emissions, 
and no He\,I lines. Fitting of the SED gives best values of T$=14000-16000$\,K and $A_V=0.4$.
At $A_V>0.45$ the temperature becomes higher than we may accept from our spectrum. 
Note \citet{Szeifert1996} used $A_V=0.8$, and their luminosity derived is 10\/\% higher.

\noindent
{\bf Var\,83}.
The star did not show strong variability in 1997 -- 2004 \citep{Direct1,Massey2006}; 
its variations were less than 0.5 mag. 
Its nebula indicates
$A_V\approx1.35$. 
In our spectrum there are very strong He\,I, Fe\,II, [Fe\,II], [Fe\,III] and Si\,II emissions
(no Bowen blend), which implies a temperature range 15000 -- 25000\,K. Our SED solutions give 
T$>30000$\,K at $A_V\approx1.35$, which is in contradiction with the spectrum. 
Hence we use two values $A_V=0.7$ (Fig.\,\ref{SED}) and 1.0 giving the most 
close temperatures to those estimated from the spectrum.
Note \citet{Szeifert1996} used $A_V=1.05$, they obtained the star luminosity 
inside the range of our limits Table\,2 
.

\noindent
{\bf Var\,2}.
The star has not shown strong variability since 1935 ($<0.3$ mag), now its 
brightness is about $V\approx18.2$ \citep{Szeifert1996,Massey2006}. 
There are no 2MASS and Spitzer data (the star is faint).
We show the K band point in Fig.\,\ref{SED} from  
\cite{Szeifert1996} as an upper limit. We found $A_V\approx0.85$ 
in Var\,2's nebula. There are He\,I, He\,II and Bowen blend lines in Var\,2 spectrum. The temperature  
is estimated as 35000 -- 50000\,K. The best fit of the SED gives estimates $A_V\approx0.8$ and 
T$\approx40000$\,K. 

\noindent
{\bf V\,532}.
The star shows strong historical variability, however in the period 1997 -- 2005 
it was relatively calm \citep[][and our own unpublished data]{Kurtev_etal2001,ZharShol04,Viotti2007},
from $B\approx17.2$ in 1997 to 17.4 in 2000, dropping to 17.7 to October 2001, and 
returning to 17.4 in 2004 -- beginning of 2005. 
A close nebula surrounding V\,532 indicates $A_V \sim 1.1$.
There are He\,II, Bowen blend and strong He\,I emissions in our spectra, 
which implies that the temperature has to be 30000 -- 40000\,K. 
The best fit of the SED we find with T\,=\,35000\,K ($A_V=0.8$). 

\begin{figure}
\includegraphics[width=\columnwidth,angle=0]{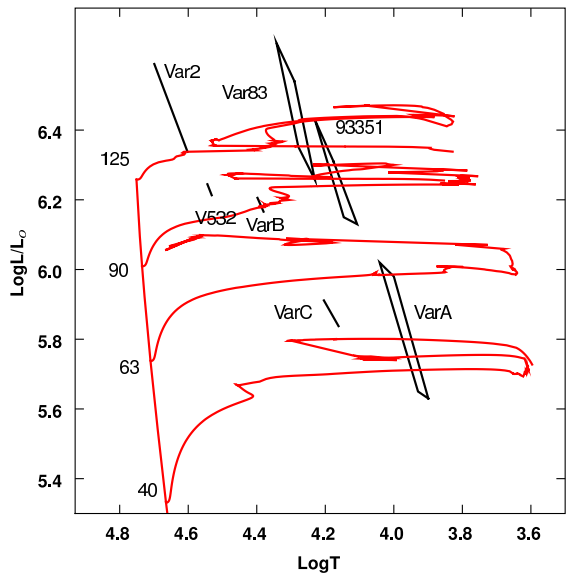}
\vspace{-0.8cm}
\caption{Temperature -- luminosity diagram for seven studied M\,33 stars. 
In N\,93351 and Var\,83 we used two most probable values of extinction, inner area shows the 
acceptable L, T values. In Var\,A our analysis is rather formal (see text).
In other stars we used one value of interstellar 
extinction (Table\,2 
). The segments indicate a range of acceptable temperatures. 
Stellar evolution tracks \citep{Claret_evol_track} for Z=0.007 stars are plotted.
}
\label{HR}
\end{figure}

\section{Concluding remarks}

Fig.\,\ref{HR} shows the temperature -- luminosity diagram for the seven studied stars 
in M\,33. Stellar evolution tracks \citep{Claret_evol_track} for 40 -- 125\,$M_\odot$
stars are plotted, which were abridged in the latest evolution stages so as not 
to crowd the figure. All these stars have masses of 
$\sim 40 - 130 \,M_\odot$, and are located in the region of evolved LBV stars. 
The new star N\,93351 is among the brightest evolved stars in M\,33. 

One finds from Table\,2 
 that the dust reradiates on average 
$\sim 8\,\%$ of the energy budget of the stars. The total scatter of stellar 
component luminosities is 1.0 dex, while that of the dust luminosities is 2.0 dex. 
Any possible uncertainties in interstellar extinction
are not as important in IR radiation as they are in the optical range, 
but we observe much bigger scatter in the dust component luminosities than that in 
stellar components. 
This may mean that the dust 
phenomenon in the most massive evolved stars is quite temporary, and that the 
dust components must be variable on scales much less than those of core
evolution. 
One may conclude that the dust activity of LBV and related stars follows 
the strong eruptions (at time scale of tens, or possibly, hundreds of years), 
as was detected in Var\,A in 1950--1953 \citep{Humph_etal1987}. 

The new star N\,93351 has properties intermediate between the LBVs and Var\,A
, in the same time it has a hot photosphere, LBV-like luminosity and an 
extensive curcumstellar material. 
This makes N\,93351 very important for detail studies in future.


\section*{Acknowledgments}

The authors thank E.\,Chentsov for helpful discussions, the referee R.\,Humphreys
for very important comments, P.\,Massey 
for providing us with M\,33 photometrical data before publication, O.\,Maryeva 
for help in spectral data reduction, P.\,Boley for correction of the manuscript.
This work is based in part on archival data obtained with the Spitzer Space 
Telescope, which is operated by the Jet Propulsion Laboratory, California 
Institute of Technology under a contract with NASA.
The research was supported by RFBR grants N\,06-02-16865, 07-02-00909.

\end{document}